\documentclass[a4paper,twocolumn]{article}
\usepackage[dvips]{graphicx}

\begin{document}

\twocolumn[
\title{Crystal Growth of Isotactic Polystyrene in Ultrathin Films : Film
 Thickness Dependence}
\author{Ken Taguchi \and Hideki Miyaji\and Kunihide Izumi\and 
Akitaka Hoshino \and Yoshihisa Miyamoto \and Ryohei Kokawa \\
Department of Physics, Graduate School of Science, \\
KyotoUniversity, Kyoto 606-8502 Japan \\
Department of Fundamental Sciences, Faculty of Integrated 
Human Studies, \\
Kyoto University, Kyoto 606-8501 Japan \\ 
Shimadzu Corporation, Kanagawa 259-1304 Japan \\}

\maketitle

\begin{abstract}
The film thickness dependence of crystal growth is investigated for	
isotactic polystyrene (it-PS) in thin films, the thickness of which is 
from 20nm down to 4nm. The single crystals of it-PS grown at 180Ž in the 
ultrathin films show the morphology typical in the diffusion-controlled 
growth: dense branching morphology (DBM), fractal seeweed (FS). The
 characteristic length of the morphology, i.e. the width of the
branch,	increases with decreasing film thickness. The thickness dependence 
of the growth rate of crystals shows a crossover around the lamellar 
thickness of the crystal, 8 nm. The thickness dependences of the growth 
rate and morphology are discussed in terms of the diffusion of chain 
molecules in thin films.
\end{abstract}
\vspace{3mm}


\vspace{1cm}
]

\section{Introduction}

Crystal growth of polymers in thin films has recently studied on	
morphology and growth rate
\cite{Izumi1997,Sawamura1998,Reiter1998,Reiter2000,Sommer2000,Taguchi2001}.
 The crystal growth rate is reduced in thin films
\cite{Frank1996,Despotopoulou1996b,Izumi1997,Sawamura1998,Reiter2000,Taguchi2001}.
For isotactic polystyrene (it-PS), the thickness dependence of the
growth rate down to 20nm was found to be expressed as 
follows ,
\begin{equation}
 G(d)=G(\infty)\left(1-a/d\right),
 \label{G-1/d}
\end{equation}
where $G(d)$ is the growth rate in the film of thickness $d$, $G(\infty)$
is the growth rate in the bulk and $a$ is a constant of about 6 nm
independent of crystallization temperature, molecular weight and
substrate materials. This decrease in growth rate was attributed to the
reduction on mobility of chain molecules in thin films and proposed that
the value of $a$ corresponds to the tube diameter in the reptation model
of polymer dynamics \cite{Sawamura1998}.

In a previous paper \cite{Taguchi2001}, we reported on morphological
change with crystallization temperature for it-PS crystals in a
ultrathin film, the thickness of which is 11nm; the crystal morphology
changes from faceted to branched one with decreasing crystallization
temperature. In the present paper we investigate the film
thickness dependence of morphology and growth rate of it-PS crystals
grown at 180$^{\circ}$C in ultrathin films, the thickness of which is
thinner than 20 nm, down to 4 nm. It is to be noted that at 180$^{\circ}$C
in the bulk it-PS the growth rate is maximum and the morphology shows a
circular disk. It is interesting to observe how the thickness dependence
of growth rate, eq (\ref{G-1/d}), together with morphology changes for
$d\leq a$. We will discuss the film thickness dependence of growth rate
and morphology in terms of chain diffusion in thin films.

\section{Experimental}

The sample used was it-PS supplied by Polymer
Laboratory ($M_{w}=590,000$, ${M_{w}}/{M_{n}}=3.4$, tacticity: 97\%
isotactic triad). This molecular weight corresponds to 22 nm in radius of
gyration of a polystyrene molecule in the melt. Ultrathin films of the it-PS
were prepared on a carbon-evaporated glass slide
by spin-coating 0.3 to 1.0wt\% cyclohexanone solution at 4000 rpm;
amorphous it-PS films with uniform thickness of 4 - 20 nm
were thereby obtained. The films were crystallized isothermally at several
temperatures 180$^{\circ}$C for a certain period of time in a hot stage
(Mettler FP800). Before crystallization, the films were melted at 250
$^{\circ}$C for 3 minutes, quenched to room temperature much lower than
the glass transition temperature Tg ($\sim$ 90$^{\circ}$C), and
immediately elevated to the crystallization temperature. The lateral
growth rate was determined by in situ 
differential-contrast optical microscopy (Nikon Optiphoto-2). Detailed
morphology and structure of the crystals were investigated by an atomic
force microscope (AFM) (SHIMADZU SPM-9500J) and a transmission electron
microscope (TEM) (JEOL 1200EX II) at room temperature after
crystallization and quenching.

\section{Results}

\begin{figure}[htbp]
 \begin{center}
  \includegraphics[width=7cm,clip]{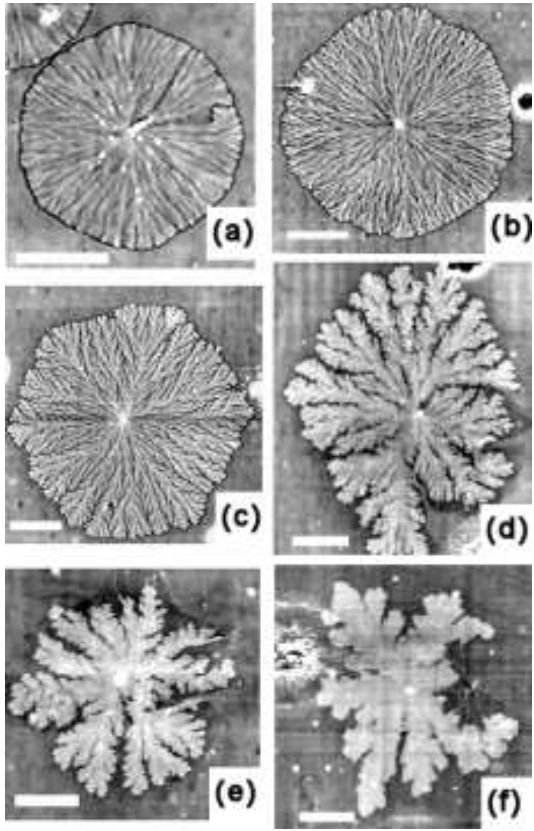}
 \end{center}
 \caption{AFM images of it-PS crystals grown at 180$^{\circ}$C in
 ultrathin  films. Each crystal is grown (a) in a film 17 nm thick for
 30min, (b) 14 nm, 1 hr, (c) 11 nm, 1 hr 30 min (d) 9.7 nm, 3 hr 15 min,
 (e) 8.7 nm, 3 hr 15min (f) 6.2 nm, 8hr 30min. Scale bars represent 5 $\mu$m.}
\label{180C_afm}
\end{figure}

Figure \ref{180C_afm} shows the AFM images (height mode) of it-PS
lamellar crystals grown at 180$^{\circ}$C in ultrathin films below 20
nm in thickness. Amorphous regions around the crystals close to the growth
interface is always thinner by several nanometers than the region far
from the interface(original thickness); we call hereafter this region
``halo'', named
after the bright region in TEM images \cite{Izumi1997}. A few nm-thick
amorphous layer covers the surface of these lamellae, including lateral
growth surface \cite{Izumi1994,Simon1996}. In 17 nm-thick films
Fig.\ref{180C_afm} a,
the crystal shows a rounded shape with an undulated growth interface and
a few small overgrowth lamellae is on its surface. It is observed that
the top surface of the lamellar crystal is also undulated a few nm in
height as indicated by the radial stripes. In films thinner than 15nm,
the crystal morphology changes to that observed in the
diffusion-controlled growth \cite{Taguchi2001}, as shown in
Fig.\ref{180C_afm} b to f. In 14 nm thick
films, the crystal shows the splitting of growth face to have many
irregular branches typical in the dense branching morphology (DBM) or
compact seaweed	(CS) \cite{Ben-Jacob1986,Brener1992} . The envelope of
this branching structure is nearly circular; we call hereafter this
envelope "average front" in contrast to the local growth interface of
its branch \cite{Brener1998}. The crystal 
grown in 11nm-thick films (Fig.\ref{180C_afm} c), also consists of many
branches, however it has an almost hexagonal average front similar to
the compact dendrite (CD)\cite{Brener1992,Brener1998}. Below about 8 nm
, the average front is triangular rather than hexagonal; it is to
be noted that it-PS has the trigonal crystal structure
\cite{Izumi1997}. There is little material left between the
branches developed.

\begin{figure}[htbp]
 \begin{center}
  \includegraphics[width=6cm,clip]{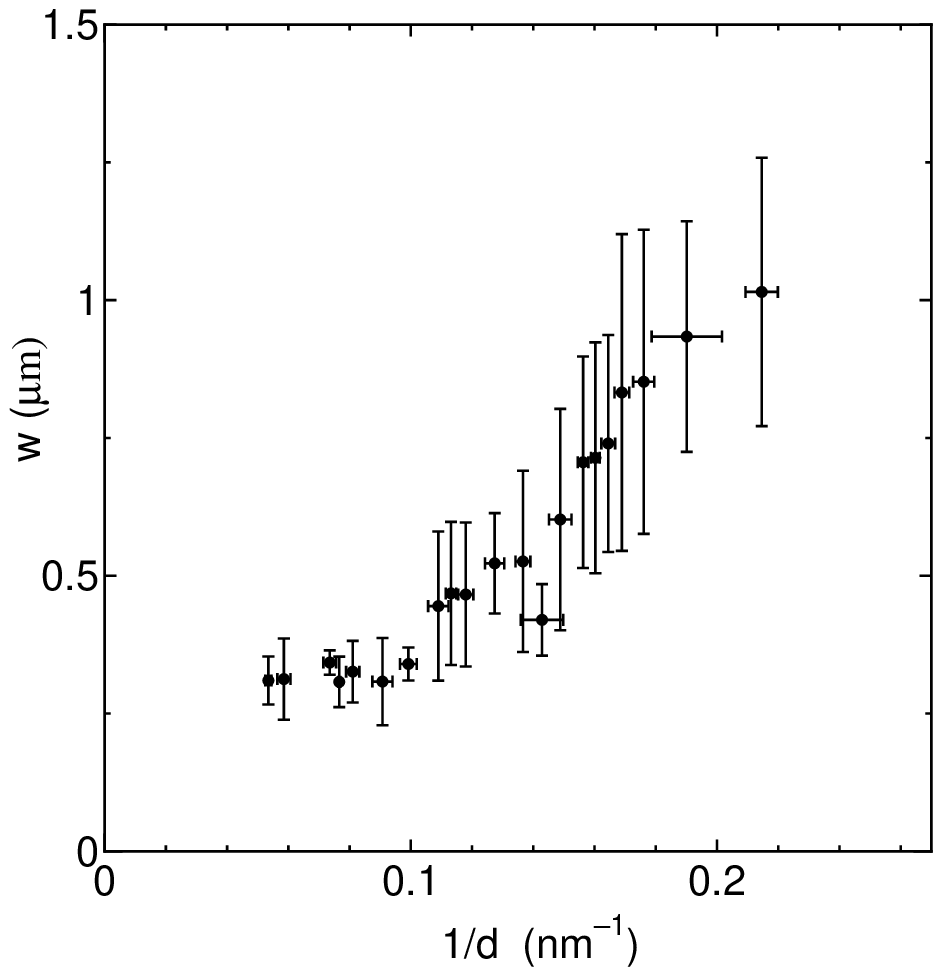}
 \caption{The mean width of branches $w$ vs. inverse of film thickness
 $1/d$ of it-PS crystals grown at 180$^{\circ}$C}
\label{180_1/d_vs_w}
 \end{center}
\end{figure}

It is clearly shown in Fig.\ref{180C_afm} d-f that for films thinner than 10nm the mean width of the branches increases as the film thickness decreases.
While the branched structures are compact in films thicker than 10nm,
those in filmsthinner than 10nm are rather open structure similar to the
fractal dendrite (FD) or seaweed (FS)
\cite{Brener1992,Brener1998}. Figure \ref{180_1/d_vs_w} shows the dependence 
of the mean width of branches, $w$, in CS or side branches in dendritic 
structure on the inverse of film thickness, $d$; the value of $w$ remains 
constant down to the film thickness of 10nm and becomes larger and larger 
with decreasing film thickness.

\begin{figure}[htbp]
 \begin{center}
  \includegraphics[width=5cm,clip]{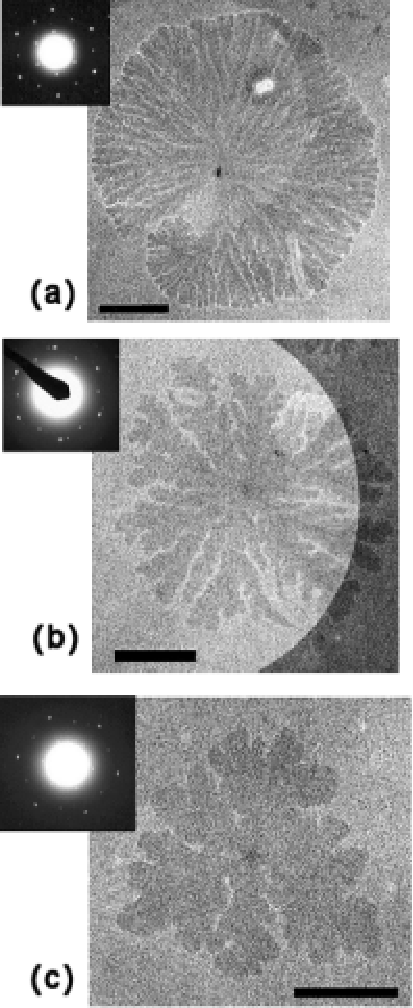}
 \caption{Electron micrographs and diffraction patterns of it-PS
 crystals grown at 180$^{\circ}$C in (a) 11 nm-thick films, (b) 8 nm and
 (c) 6nm respectively. Scale bars represent 2$\mu$m}
 \label{180C_tems}
\end{center}
\end{figure}

Although the top surface of the lamellar crystal grown in films 
thicker	than 11 nm is at almost the same level as that of surrounding melt, 
the crystals in films thinner than 10 nm protrudes from the surrounding 
amorphous surface. The regions of ``halo'' extend over a few hundreds nm 
radially in the	thicker films and are observed in the AFM images as dark 
contrast in front of the growth face of a crystal. In films thinner than 
10 nm, the width of the ``halo'' becomes larger with decreasing film 
thickness and eventually no ``halo'' is observed. Figure \ref{180C_tems}
shows the TEM bright field images of it-PS crystals grown in ultrathin
films at 180$^{\circ}$C with the diffraction patterns. The diffraction
patterns prove that all the branching crystals have a single
crystallographic orientation with the chain axis perpendicular to the
lamellar surface. It is also found that the three sides of the
triangular crystals are parallel to the $\{1\bar{2}10\}$ planes.

\begin{figure}[htbp]
 \begin{center}
  \includegraphics[width=6cm,clip]{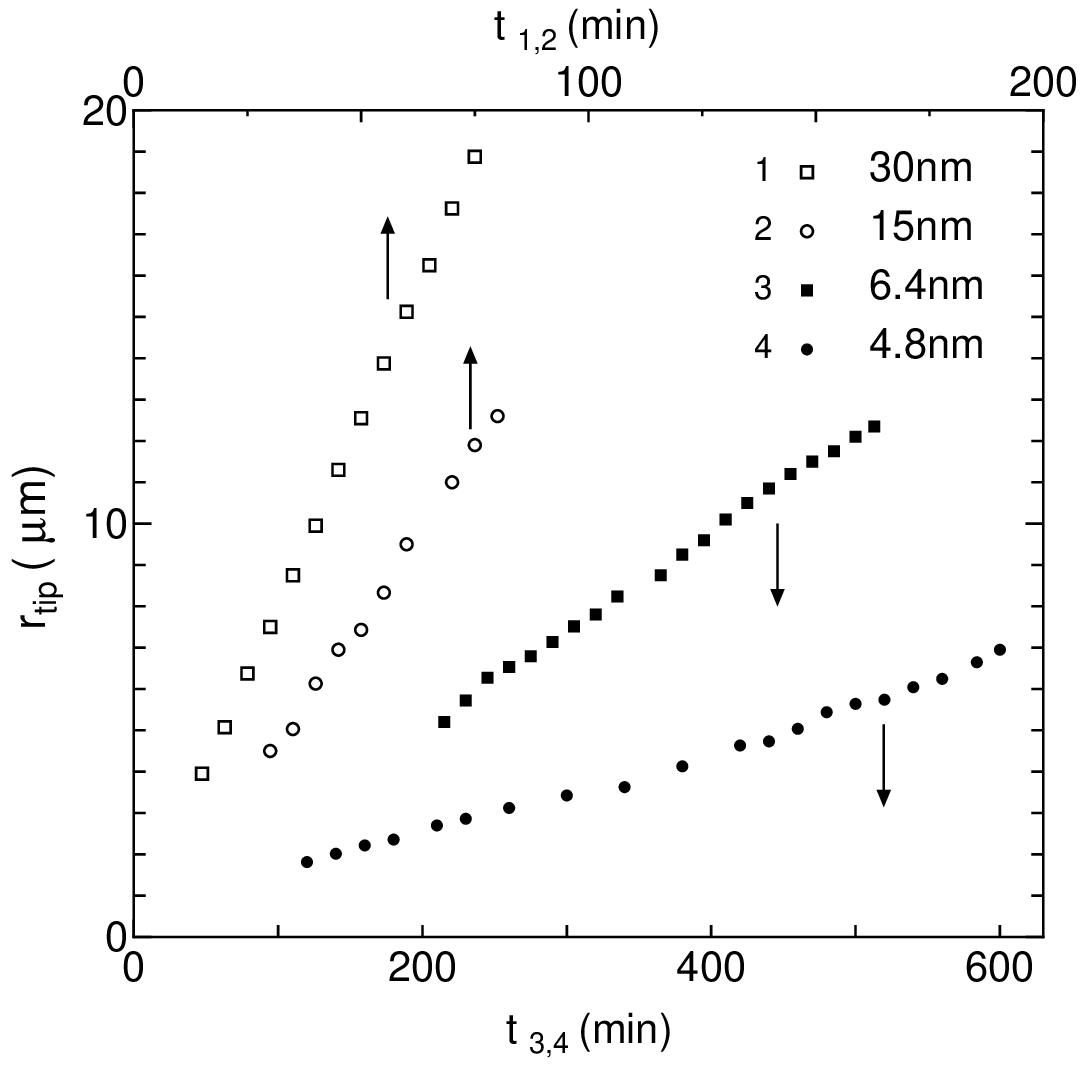}
 \caption{Time evolutions of $r_{tip}$ of it-PS crystals grown at
 180$^{\circ}$C in each film thickness. The upper abscissa is for the
 data of 30 nm and 15 nm thick films and the lower abscissa for 6.4 nm
 and 4.8 nm.}
\label{180C_in-situ}
 \end{center}
 \end{figure}

\begin{figure}[htbp]
 \begin{center}
  \includegraphics[width=6cm,clip]{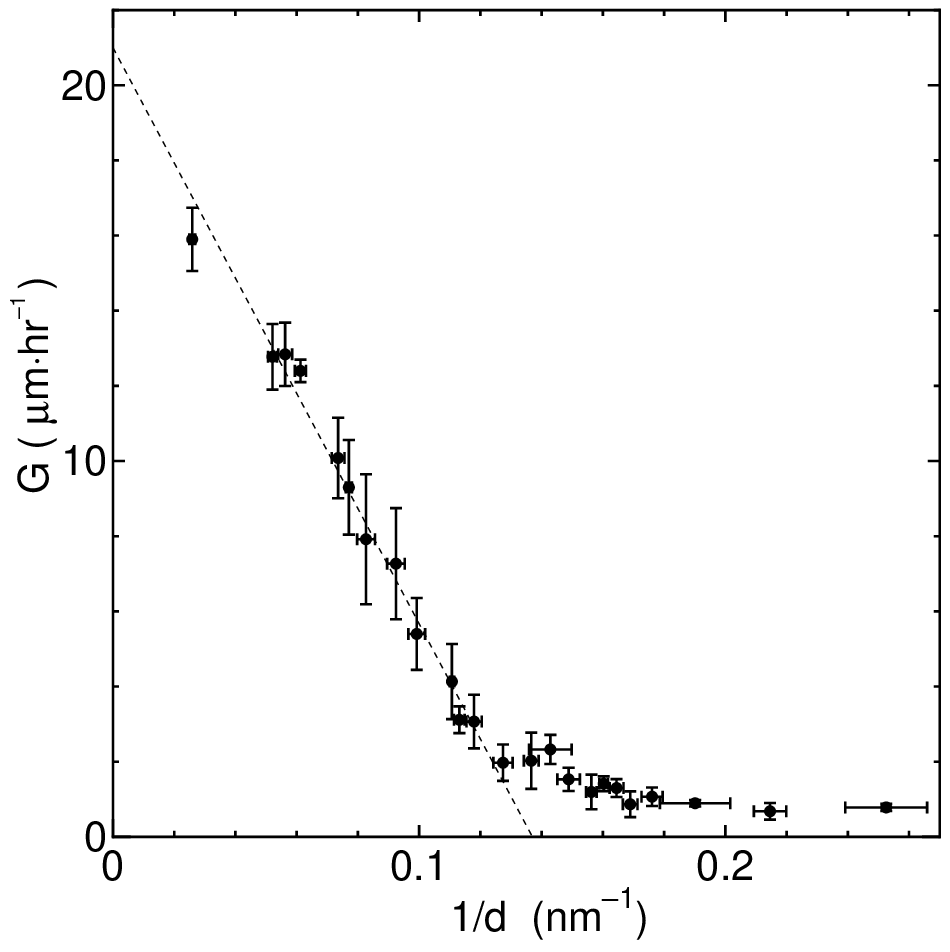}
 \caption{Growth rate $G(d)$ vs. inverse of film thickness
  $1/d$ for it-PS crystals grown at 180$^{\circ}$C in ultrathin
  films. The dotted line shows $G(d)=G(\infty)(1-a/d)$ with fitting
  parameters $a=$7.2 nm and $G(\infty)=$ 21 $\mu$m/hr.}
 \label{180_1/d_vs_G}
\end{center}
\end{figure}

Figure \ref{180C_in-situ} shows the time evolution of the farthest
tips from the center of crystals, $r_{tip}$, measured by in-situ optical
microscopy.
All the growth rates of farthest tips in ultrathin films down to 5
nm-thick were found to be constant during growth (linear growth); it is
to be noted that the growth rates of reentrant sites between the
fartheset tips decreased during growth. The growth rates of farthest tips
determined from the data in Fig. \ref{180C_in-situ} are plotted against
the inverse of film thickness, $1/d$, as shown in
Fig. \ref{180_1/d_vs_G}. The growth rates of it-PS grown at
180$^{\circ}$C in ultrathin films also decrease with decreasing film
thickness according to the Eq. (\ref{G-1/d}) with $a \sim$ 7.2
nm. However below about 9 nm, the Eq. (\ref{G-1/d}) can not hold any
longer and a crossover of the film thickness dependence is observed;
the decreasing rate of $G$ with 1/$d$ becomes much lower,and in 5
nm-thick films the growth rate reduces to 1/20 of that in the bulk. It
is very interesting that the thickness of 9 nm is that of crystal
lamellar thickness grown at 180$^{\circ}$C.

\begin{figure}[htbp]
 \begin{center}
  \includegraphics[width=7cm,clip]{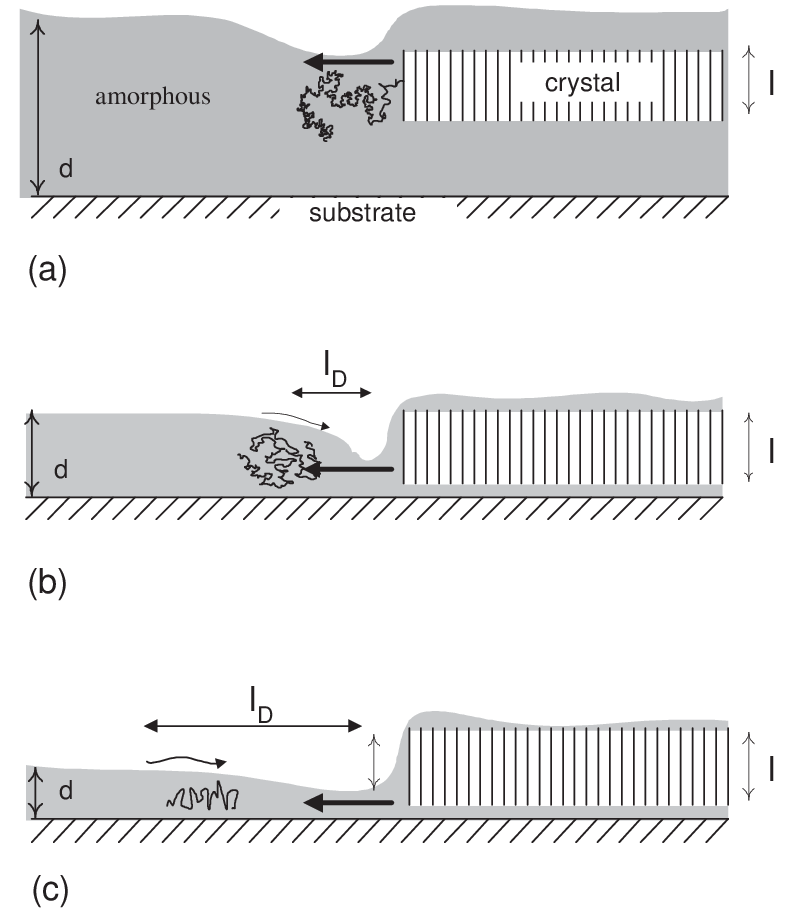}
 \end{center}
 \caption{Schematic views of the crystal growth in thin films;
 cross-sections at growth tips with the vertical scale magnified about
 100 times compared with the horizontal scale: (a) $d > l$, where $l$ is
 lamellar thickness, (b) $d \sim l$, (C) $d < l$ }
\label{thick_change}
\end{figure}

\section{Discussion}

The experimental results on the morphology and growth rates indicate
that the diffusion process of chain molecules begins to control the
crystal growth in ultrathin films \cite{Saito1996}. The morphological
instability in ultrathin films was considered to be caused by the
gradients of film thickness in the ``halo'' region, which could makes a
self diffusion field of polymer molecules \cite{Taguchi2001}. When the
film thickness is much thicker than the depletion of ``halo''of several
nanometers, the effect of the gradient on instability is so small that
the crystal remains a circular disk as schematically shown in
Fig. \ref{thick_change} a. However, when the film thickness becomes
thinner than the lamellar thickness ($\sim$ 9 nm), the growth of
crystal requires diffusion of molecules in the melt further from growth
front in order to supply crystallizing materials for the thicker crystals
(Fig. \ref{thick_change} b,c). Hence, the growth rate in films thinner than
lamellar thickness also decreases with decreasing film thickness. This
film thickness dependence of the growth rates is different from that in
films thicker than lamellar thickness caused by reduction of chain
mobility in thin films due to entanglement effect \cite{Sawamura1998},
and hence the crossover is observed at lamellar thickness. A
morphological instability appears by the gradient of film thickness
around a crystal; the characteristic length of the diffusion controlled
growth, such as the width of branches, is generally scaled by the
diffusion length $l_{D}$ given by $2D/G(d)$ where $D$ is the diffusion
constant of chain molecules \cite{KP1963,Saito1989,Taguchi2001}. For the
film thinner than the entanglement distance, i.e. the diameter of the
tube in the reptation model ($\sim$ 7 nm \cite{PPPH}),
entanglement effect on diffusion constant is small; the diffusion
constant $D$ should remain almost constant with film thickness
\cite{Reiter1998}. Still the $G(d)$ decreases with decreasing
film thickness. Consequently $l_{D}$ increases as denoted in
Fig. \ref{thick_change} b,c and hence the width of branch $w$ accordingly
increases with decreasing film thickness. When $l_{D}$ becomes very large
and then ``halo'' is hardly observed as in Fig. \ref{180C_afm} f in very
thin films, the situation looks similar to the DLA growth: the crystal
structure hence should become open fractal aggregates. In fact,
this has been demonstrated in the Monte Carlo simulation
\cite{Reiter1998,Sommer2000} on crystallization of adsorbed polymer
monolayers, where the thicker the lamellar thickness becomes, the wider
of the width of branches and more open the structure becomes. It is
worth noting that the dendritic tip can grow steadily even in the diffusion
controlled growth as in Fig. \ref{180C_in-situ}, while the steady growth
of a spherical interface is impossible; its growth rate decreases with
time as $G\propto t^{-1/2}$ \cite{Saito1989,Saito1996}.

\section{Conclusion}

In this paper we have reported the film thickness dependence of morphology
and growth rates of it-PS crystals grown at 180$^{\circ}$C in ultrathin
films thinner than 20 nm down to 4 nm. It is clearly shown that in
films thinner than 15 nm, the crystal morphology typical in the
diffusion-controlled growth appeared and the morphology varied with
decreasing film thickness from dense branching morphology (DBM) to
diffusion-limited aggregates (DLA). In particular, it is 
found that in films thinner than the lamellar crystal, the structure
becomes more open and characteristic length of the strucrure becomes
larger and larger with decreasing film thickness. The dependence of the
growth rates on film thickness also changed in the vicinity of the
lamellar thickness.

\section{Acknowledgements}

This work was supported partly by Grant-in-Aid for Science Research on
PriorityAreas, "Mechanism of Polymer Crystallization" (No12127204) 
from The Ministry of Education, Science, Sports and Culture.

\newpage

\newpage

\newpage

\newpage


\begin{thebibliography}{10}
\expandafter\ifx\csname url\endcsname\relax
  \def\url#1{\texttt{#1}}\fi
\expandafter\ifx\csname urlprefix\endcsname\relax\def\urlprefix{URL }\fi

\bibitem{Izumi1997}
Izumi,~K.;Gan,~P.;Hashimoto,~M.;Toda,~A.;
	Miyaji,~H.;Miyamoto,~Y.;Yoshitsugu,~N.
	Crystal growth of polymers in thin films, {\it In Advances in
	the Understanding of Crystal Growth Mechanisms};Nishinaga,~T.,
	Nishioka,~K., Harada,~J. Sakai,~A., Takei,~H., Eds., Elsevier
	Science, 1997, pp. 337--348.

\bibitem{Sawamura1998}
Sawamura,~S.;Miyaji,~H.;Izumi,~K.;Sutton,~S.J.; Miyamoto,~Y. Growth rate
	of isotactic polystyrene crystals in thin films.
	J. Phys. Jpn. {\bf 1998},67~(10),3338--3344.

\bibitem{Reiter1998}
Reiter,~G.;Sommer,~J.-U. Crystallization of adosrbed polymer
	monolayers, Phys. Rev. Let. {\bf 1998}, 80~(17), 3771--3774.

\bibitem{Reiter2000}
Reiter,~G.;Sommer,~J.-U. Polymer crystallization in quasi-two
	dimensions. I. Experimental Results, J. Chem. Phys. {\bf 2000},
	112~(9), 4377--4383.

\bibitem{Sommer2000}
Sommer,~J.-U.;Reiter,~G. Polymer crystallization in quasi-two
	dimensions. ii. kinetic models and computer simulations,
	J. Chem. Phys. {\bf 2000}, 112~(9),  4384--93.

\bibitem{Taguchi2001}
Taguchi,~K.;Miyaji,~H.;Izumi,~K.;Hoshino,~A.; Miyamoto,~Y.;Kokawa,~R.
	Growth shape of isotactic polystyrene crystals in thin films,
	Polymer {\bf 2001},42~(17), 7743--47.

\bibitem{Frank1996}
Frank,~C.;Rao,~V.;Despotopoulou,~M.~M.; Pease,~R.~F.~W.;Hisenberg,~W.~D.;
	Miller,~R.~D.;Rabolt,~J.~F.Structure in thin and ultrathin
	spin-cast polymer films, Science {\bf 1996}, 273, 912--915.

\bibitem{Despotopoulou1996b}
Despotopoulou,~M.~M.; Miller,~R.~D.; Rabolt,~J.~F.;Frank,~C. Polymer
	chain organization and orientation in ultrathin films: a
	spectroscopic investigation, J. Polymer. Sci. B, Polym. Phys. {\bf 
	1996}, 34~{14}, 2335--2349.

\bibitem{Izumi1994}
Izumi,~K.;Gan,~P.;Toda,~A.;Miyaji,~H.; Hashimoto,~M.;Miyamoto,~Y.;Nakagawa,~Y.
	Atomic force microscipy of isotactic polystyrene crystals,
	J. App. Phys. {\bf 1994}, 33, 1628--1630.

\bibitem{Simon1996}
Sutton,~S.~J.;Izumi,~K.;Miyaji,~H.;Fukao,~K.; Miyamoto,~Y. The lamellar
	thickness of melt crystallized isotactic polystyrene as
	determined by atomic foece microscopy, Polymer {\bf 1996}, 37,
	5529--5532. 

\bibitem{Ben-Jacob1986}
Ben-Jacob,~E.;Deutscher,~G.;Garik,~P.;
	Goldenfeld,~N.~D.;Lareah,~Y. Formation of
  a dense branching morphology in interfacial
	growth. Phys. Rev. Lett. {\bf 1986}, 57~(15), 1903-06.

\bibitem{Brener1992}
Brener,~E.;M{\"u}ller-Krumbhaar,~H.;Temkin,~D. Kinetic phase diagram and
	scailing relations for stationary diffusion growth,
	Europhys. Lett. {\bf 1992},
	17~(6), 535--540.

\bibitem{Brener1998}
Brener,~E.; M{\"u}ller-Krumbhaar,~H.; Temkin,~D.;Abel,~T. Morphology
	diagram of possible structures in diffusional growth, Physica A
	{\bf 1998},249~(1-4) 73--81.

\bibitem{Saito1996}
Saito,~Y. Statistical physics of crystal growth, World Scientific
	Publishing Co. Pte. Ltd, 1996.

\bibitem{KP1963}
Keith,~H.~D.;Padden,~F.~J.~Jr. A phenomenological theory of spherulitic
	crystallization, J. Appl. Phys. {\bf 1963}, 34, 2409--2421.

\bibitem{Saito1989}
Saito,~Y.;Ueta,~T. Monte carlo studies of equilibrium and growth shapes
	of a crystal, Phy. Rev. {\bf 1963}, A 34, 3408--3419.

\bibitem{PPPH}
Fetter,~L.J.; Lohse,~D.J.;Colby,~R.H. Chain Dimensions and Entanglement
	Spacings. In {\it Physical Properties of Polymers Handbook};
	Mark,~James E, Eds.:Woodbury, New York, American Institute of
	Physics,{\bf 1996};Ch.~24, p 335--340.


\end{thebibliography}
\end{document}